\documentclass[twocolumn,superscriptaddress,showpacs,showkeys,aps,prb]{revtex4-1}
\usepackage{amsmath}
\usepackage{amssymb}

\newlength {\defaultparindent }
\newenvironment {annotation Text}{}{}

\begin{document}

\title{The Coherent Forward Scattering Amplitude in Transmission and Grazing 
Incidence M\"{o}ssbauer Spectroscopy}
\author{L. De\'{a}k}
\email{deak.laszlo@wigner.mta.hu}
\affiliation{Wigner\ RCP, RMKI, P.O.B. 49, 1525 Budapest, Hungary}
\author{L. Botty\'{a}n}
\affiliation{Wigner\ RCP, RMKI, P.O.B. 49, 1525 Budapest, Hungary}
\author{D. L. Nagy}
\affiliation{Wigner\ RCP, RMKI, P.O.B. 49, 1525 Budapest, Hungary}
\author{H. Spiering}
\affiliation{%
Johannes Gutenberg
Universit\"{a}t Mainz, Staudinger Weg 9, 55099 Mainz, Germany}
\date{Submitted to Physical Review B on July 25, 1995}

\begin{abstract}
The theory of both transmission and grazing incidence M\"{o}ssbauer
spectroscopy is re-analyzed. Starting with the nuclear susceptibility tensor
a common concise first order perturbation formulation is given by
introducing the forward scattering amplitude into an anisotropic optical
scheme. Formulae of Blume and Kistner as well as those of Andreeva are
re-derived for the forward scattering and grazing incidence geometries,
respectively. Limitations of several previously intuitively introduced
approximations are pointed out. The grazing incidence integral propagation
matrices are written in a form built up from $2\times 2$ matrix exponentials
which is particularly suitable for numerical calculations and practical
fitting of both energy domain (conventional source experiment) and time
domain (synchrotron radiation experiment) M\"{o}ssbauer spectra.
 \end{abstract}

\pacs{PACS: 42.25-p,68.35.-p,76.80.+y,78.66.-w}
\maketitle
\preprint{HEP/123-qed}

\section{INTRODUCTION}

A great majority of M\"{o}ssbauer experiments is performed on
polycrystalline samples without applying an external magnetic field. In such
cases, the polarization of the $\gamma $-rays plays no role, the
M\"{o}ssbauer spectrum can be described in terms of resonant and
non-resonant absorption and the resonant absorption cross-section can be
calculated from the parameters of the hyperfine interaction. This naive
approach fails if the M\"{o}ssbauer experiment is performed on a single
crystal or a textured sample and/or in an external magnetic field. The
resonant cross-section in these latter cases depends on the polarization and
the full polarization-dependent scattering problem has to be treated. The
numerical difficulties of the scattering approach stem from the great number
of randomly distributed scattering centres. These difficulties can be
circumvented if, akin to classical optics, a continuum model rather than a
microscopic scattering theory can be used. It is by no means trivial,
however, that such an optical approach for $\gamma $-rays in condensed
matter is feasible since the mean distance of scattering centres is usually
greater than the wavelength of the scattered radiation. It has been shown,
however, by Lax\cite{Lax} that, at least for scalar waves, a close to unity
index of refraction $n$ can be defined and simply related to the coherent
forward scattering length $f$, provided that the momentum of the scatterers
is small compared to that of the incident wave. Since Lax's paper,\cite{Lax}
the refraction index approach has been used extensively in neutron and x-ray
optics. The heuristic generalization of this approach to polarized waves and
for an anisotropic medium, although claimed to be trivial by Lax is by no
means straightforward and needs further elucidation.

In the forward scattering geometry the polarization dependence of the
M\"{o}ssbauer absorption of $\gamma $-radiation was theoretically studied by
Blume and Kistner.\cite{Blume} Instead of using a $3\times 3$ index of
refraction matrix, they accepted Lax's intuitive suggestion,\cite{Lax} and
used a complex $2\times 2$ index of refraction matrix $n$, corresponding to
the two possible independent states of polarization of the radiation. $n$
was then related to the coherent forward scattering amplitude.\cite{Blume}

Beside the conventional forward scattering case, grazing incidence
M\"{o}ssbauer spectroscopy (GIMS) has gained considerable recent attention
in studying stratified media: surfaces, interfaces and multilayers.\cite
{Frost,Nagy,Irkaev1,Irkaev2,Andreeva2} This method utilizes a geometry such
that the $\gamma $-rays are incident on the flat surface of the sample at
glancing angles of a few mrad close to the critical angle of the electronic
total external reflection. The detected scattered particles are specularly
reflected $\gamma $-photons, conversion electrons, conversion x-rays and
incoherently scattered $\gamma $-photons. A general treatment of GIMS was
published by Andreeva {\it et al.} in several papers.\cite
{Irkaev2,Andreeva2,Andreeva1,Andreeva3} Starting from the nucleon current
density expression of the susceptibility tensor $\chi $ given by Afanas'ev
and Kagan\cite{Kagan} and using a covariant formalism of anisotropic optics%
\cite{Borzdov} first introduced by Fedorov\cite{Fedorov} these authors take
into account that both the elements of the susceptibility tensor $\chi $ and
the glancing angle $\theta $ are small in GIMS and calculate the $\gamma $%
-reflectivity. The method of calculation, however, especially for the higher
multipolarity nuclear transitions, is rather cumbersome, since the nucleon
current densities are directly calculated resulting in quite complex tensor
expressions. In view of the extreme requirements to beam divergence, GIMS is
certainly more suited for synchrotron radiation than for conventional
radioactive source experiments.

Another general description of specular reflection of grazing incidence
M\"{o}ssbauer radiation was given by Hannon {\it et al.}\cite
{Hannon1,Hannon2,Hannon3,Hannon4} Starting from the quantum theory of $%
\gamma $-radiation, they formulated the dynamic theory of M\"{o}ssbauer
optics. Unfortunately, the dynamic theory provides rather slow algorithms
for calculating reflectivity spectra, therefore it is inefficient in
spectrum fitting. In the grazing incidence limit, an optical model was
derived from the dynamical theory,\cite{Hannon2,Hannon4} which has recently
been implemented in numerical calculations.\cite{Roehlsberger} Without using
a covariant formalism, however, this latter approach also results in quite
sophisticated algorithms since, in a layered medium, the eigenpolarizations
vary from layer to layer.

Our aim is to rigorously derive general formulae for the transmissivity and
the reflectivity of $\gamma $-radiation in the forward scattering and the
grazing incidence case, respectively. Moreover, we shall try to obtain these
formulae in such form that is suitable for fast numerical calculations in
order to {\it fit }the experimental data. Like Andreeva {\it et al.},\cite
{Irkaev2,Andreeva2,Andreeva1,Andreeva3} we start from the Afanas'ev--Kagan
nucleon current density expression of the dielectric tensor\cite{Kagan} and
use a covariant anisotropic optical formalism.\cite{Borzdov,Fedorov} Instead
of calculating the susceptibility tensor from the current densities of the
nucleons, however, we reduce the problem to the calculation of the
transmittance (forward scattering case) and the reflectivity (grazing
incidence case) from the coherent forward scattering amplitude. We show
that, in the case of forward scattering, this approach is equivalent to the
theory of Blume and Kistner.\cite{Blume} The present treatment is based on
no intuitive assumption and represents, thereby, a firm basis of the
Blume--Kistner theory\cite{Blume} and of the Andreeva approximation\cite
{Irkaev2,Andreeva2,Andreeva1,Andreeva3}.

\section{THE NUCLEAR SUSCEPTIBILITY}

Let us consider the collective system of the nuclei and the electromagnetic
field. The effect of the electromagnetic field will be treated as a
perturbation on the randomly distributed nuclei. The interaction Hamiltonian 
$H$ between the nucleus and the electromagnetic field may be written as 
\begin{equation}
H=-\frac 1c\sum\limits_i{\bf j}({\bf r}_i)\cdot {\bf A}({\bf r}_i)\text{,}
\label{Hamilton}
\end{equation}
where ${\bf j}({\bf r}_i)$ is the current density of the $i$-th nucleon and $%
{\bf A}({\bf r}_i)$ is the vector potential of the electromagnetic field at
the point ${\bf r}_i$:
\begin{equation}
{\bf A}({\bf r}_i)=\sum\limits_{k,p}\left( \frac{2\pi \/\hbar c}{Vk}\right)
^{1/2}\left\{ c_{{\bf k},p}\widehat{{\bf u}}_{{\bf k},p}\exp \left( i{\bf %
k\cdot r}_i\right) +\text{H.C.}\right\} \text{.}  \label{Photon}
\end{equation}
In this formula, $c_{{\bf k},p}$ denotes a photon annihilation operator and $%
\widehat{{\bf u}}_{{\bf k},p}$ a unit polarization vector.

The matrix elements of the interaction Hamiltonian $H$ are scalar products
of the current density matrix elements ${\bf J}_{m_gm_e}$ and the
polarization vectors $\widehat{{\bf u}}_{{\bf k},p}$ :
\begin{widetext} 
\begin{eqnarray}
H_{m_gm_e}^{{\bf k},p} &=&\left( \frac{2\pi \hbar c}{Vk}\right)
^{1/2}\left\langle I_gm_g{\bf k}p\left| \sum\limits_i\left( -\frac 1c\right) 
{\bf j}\left( {\bf r}_i\right) \cdot \widehat{{\bf u}}_{{\bf k},p}c_{{\bf k}%
,p}^{\dagger }\exp \left( -i{\bf k\cdot r}_i\right) \right|
I_em_e\right\rangle  \nonumber  \\
&=&-\frac 1c\left( \frac{2\pi \hbar c}{Vk}\right) ^{1/2}\widehat{{\bf u}}_{%
{\bf k},p}\cdot {\bf J}_{m_gm_e}  \label{Jdef}
\end{eqnarray}
\end{widetext} with 
\begin{equation}
{\bf J}_{m_gm_e}\left( {\bf k}\right) =\left\langle I_gm_g\left|
\sum\limits_i{\bf j}\left( {\bf r}_i\right) \exp \left( -i{\bf k\cdot r}%
_i\right) \right| I_em_e\right\rangle  \label{Jmgme}
\end{equation}
where $I_g$ and $I_e$ are the nuclear spin quantum numbers in the excited
and ground state with the corresponding magnetic quantum numbers $m_g$ and $%
m_e$, respectively. ${\bf J}({\bf k})$ is the ${\bf k}$-representation of
the current density produced by a single nucleus. (Throughout the
calculations we shall use the same letters for physical quantities in ${\bf r%
}$- and ${\bf k}$-representation letting the argument make evident which
representation is meant.)

In first order perturbation of the electromagnetic field the average nucleon
current density reads\cite{Kagan}
\begin{equation}
\left\langle {\bf J}^{(1)}\left( {\bf k}\right) \right\rangle =\sigma \left( 
{\bf k}\right) \,{\bf E}\left( {\bf k}\right) =\frac{i\omega }c\sigma \left( 
{\bf k}\right) {\bf \,A}\left( {\bf k}\right) \text{.}  \label{Current}
\end{equation}
$\,$with $\sigma \left( {\bf k}\right) \,$ being the conductivity tensor,
which in turn defines the susceptibility tensor of the medium by
\begin{equation}
\chi \left( {\bf k}\right) =\frac{4\pi \,i}\omega \sigma \left( {\bf k}%
\right) \text{.}  \label{EpszM}
\end{equation}

Afanas'ev and Kagan\cite{Kagan} calculated the susceptibility tensor in
first order of the vector potential for randomly distributed nuclei in terms
of the change of the average nucleon current density:
\begin{equation}
\chi \left( {\bf k}\right) =-\frac{4\pi }{c^2k^2}\frac N{2I_g{}+1}%
\sum\limits_{m_em_g}\frac{{\bf J}_{m_gm_e}{}\left( {\bf k}\right) \circ {\bf %
J}_{m_em_g}^{*}{}\left( {\bf k}\right) }{E_{{\bf k}}-E_{m_em_g}+\frac{%
i\Gamma }2}\text{,}  \label{EpszK}
\end{equation}
where $N$ is the number of resonant nuclei per unit volume, $E_{{\bf k}}$ is
the energy of the $\gamma $-photon, $E_{m_em_g}=E_{m_e}-E_{m_g}$ is the
energy difference between the nuclear excited and ground states, $\Gamma $
is the natural width of the excited state and $\circ $ is the dyadic vector
product sign. The susceptibility tensor $\chi \left( {\bf k}\right) $
depends on the propagation vector ${\bf k}$ of the unperturbed wave. Instead
of Eq. (\ref{Current}) a $\left\langle {\bf J}^{(1)}\left( {\bf k}\right)
\right\rangle =\sum\limits_{{\bf K}}\sigma \left( {\bf k,k+K}\right) \,{\bf E%
}\left( {\bf k+K}\right) $ expression is obtained for non-random
distribution of the scatterers,\cite{Kagan} with $\left( 2\pi \right) ^{-1}%
{\bf K}$ being a reciprocal lattice vector. Only the random scatterer case
will be further considered here.

Eq. (\ref{EpszK}) is the starting equation of Andreeva in calculating
grazing incidence M\"{o}ssbauer spectra.\cite{Andreeva1} In order to
calculate $\chi \left( {\bf k}\right) $ for an arbitrary orientation of the
hyperfine fields with respect to ${\bf k}${\bf ,} the currents ${\bf J}%
_{m_gm_e}$ are expanded in terms of irreducible tensors\cite{Andreeva1}. For
cases, like transitions of higher multipolarity, mixed multipole
transitions, variation of hyperfine fields within the medium, texture, {\it %
etc.} the formalism therefore becomes cumbersome and numerically
intractable. Having calculated the dielectric tensor of the M\"{o}ssbauer
medium, Andreeva {\it et al.} apply a very elegant covariant formalism\cite
{Borzdov} and solve the problem of grazing incidence nuclear scattering by
stratified media.

The numerical difficulties of the higher multipolarity terms, hyperfine
field distributions, texture, {\it etc}{.} have been overcome years ago by
Spiering\cite{Spiering85} in treating the thick absorber case in the
Blume--Kistner formalism.\cite{Blume} The Hamiltonian, the scalar product of
the current density ${\bf J}_{m_gm_e}$ and the polarization vector $\widehat{%
{\bf u}}_{{\bf k},p}$ have simpler transformation properties than ${\bf J}%
_{m_gm_e}$, therefore, unlike Andreeva {\it et al.},\cite{Andreeva1} the
forward scattering amplitude
\begin{equation}
f^{{\bf k},p\rightarrow {\bf k},p^{^{\prime }}}=-\frac{kV}{2\pi \;\hbar c}%
\frac 1{2I_g{}+1}\sum\limits_{m_em_g}\frac{H_{m_gm_e}^{{\bf k},p^{^{\prime
}}}H_{m_em_g}^{{\bf k},p\dagger }{}}{E_{{\bf k}}-E_{m_em_g}+\frac{i\Gamma }2}%
\text{.}  \label{f}
\end{equation}
rather than ${\bf J}_{m_gm_e}$ is calculated for an arbitrary ${\bf k}$%
-direction.\cite{Spiering85}

In what follows we shall show that for $\gamma $-quanta in the physically
relevant representation the $3\times 3$ properties of the dielectric tensor
are not fully used by the optical theory. Since the ${\bf k}$-directions
involved in the scattering problem are either equivalent (forward
scattering) or extremely close to each other (grazing incidence case), the
relevant block of the dielectric tensor is fully described by the four
components of the forward scattering amplitude. This latter ensures that the
present theory remains valid for nuclear transitions of any multipolarity.
Indeed, expressing the susceptibility tensor in the polarization vector
system ${\cal P=\,}\left( \widehat{{\bf e}}_{\sigma ,\pi }=\widehat{{\bf u}}%
_{\sigma ,\pi }\,,\,\widehat{{\bf e}}_3={\bf k/}\left| {\bf k}\right|
\right) $ of the unperturbed incident radiation the significant matrix
elements are:
\begin{equation}
\chi _{pp^{^{\prime }}}\left( {\bf k}\right) =\frac{4\pi N}{k^2}f^{{\bf k}%
,p\rightarrow {\bf k},p^{^{\prime }}}\quad p,\ p^{^{\prime }}=\sigma ,\pi 
\text{;}  \label{fDeak}
\end{equation}
$\sigma $ and $\pi $ being arbitrary polarizations. Once the susceptibility
(the refractive index or the dielectric) tensor of the medium is defined the
problem of calculating the propagation of electromagnetic field in the
medium becomes an optical problem. Since the nuclear dielectrics is
anisotropic, a polarization-dependent optical formalism will be used.

\section{Covariant anisotropic optics of a nuclear dielectrics}
\label{3}

The covariant optical formalism of stratified anisotropic media developed by
Borzdov, Barskovskii and Lavrukovich\cite{Borzdov} and applied by Andreeva 
{\it et al.}\cite{Irkaev2,Andreeva2,Andreeva1,Andreeva3} will be introduced
here for three reasons:
\begin{enumerate}
\item  Approximations made by Andreeva {\it et al.} are based on the
assumption that the square of the scattering angle is of the order of the
susceptibility tensor elements. The borderline of the Andreeva approximation
will be specified here.

\item  The Blume--Kistner theory\cite{Blume} will be derived from the
covariant optical formalism.

\item  In a practical application of the Blume--Kistner theory one
calculates the exponentials of $2\times 2$ complex matrices. The covariant
optics uses $4\times 4$ matrices in the exponentials leading to rather time
consuming calculations. It will be shown that in a suitably chosen basis,
the $4\times 4$ matrices reduce to $2\times 2$ ones both in forward
scattering and in grazing incidence geometry.
\end{enumerate}

\subsection{The Borzdov--Barskovskii--Lavrukovich formalism}

We may write the basic equation for the tangential components of the
electric and magnetic fields $\widehat{{\bf q}}\times {\bf E}\left( \widehat{%
{\bf q}}\cdot {\bf r}\right) $ and ${\bf H}_t\left( \widehat{{\bf q}}\cdot 
{\bf r}\right) =-\widehat{{\bf q}}\times \left[ \widehat{{\bf q}}\times {\bf %
H}\left( \widehat{{\bf q}}\cdot {\bf r}\right) \right] $ at the point ${\bf r%
}$ as follows:\cite{Borzdov}
\begin{equation}
\left( \widehat{{\bf q}}\cdot \nabla \right) \left( 
\begin{array}{c}
{\bf H}_t\left( \widehat{{\bf q}}\cdot {\bf r}\right) \\ 
\widehat{{\bf q}}\times {\bf E}\left( \widehat{{\bf q}}\cdot {\bf r}\right)
\end{array}
\right) =ikM\left( \widehat{{\bf q}}\cdot {\bf r}\right) \left( 
\begin{array}{c}
{\bf H}_t\left( \widehat{{\bf q}}\cdot {\bf r}\right) \\ 
\widehat{{\bf q}}\times {\bf E}\left( \widehat{{\bf q}}\cdot {\bf r}\right)
\end{array}
\right) \text{,}  \label{BorzEq}
\end{equation}
were $\widehat{{\bf q}}$ represents the unit normal vector of the surface.
The material parameters are allowed to vary only in the $\widehat{{\bf q}}$%
-direction (stratified medium) and the fields depend only on the $\widehat{%
{\bf q}}\cdot {\bf r}$ scalar product. $M$ is the differential propagation
matrix defined by
\begin{equation}
M=\left( 
\begin{array}{cc}
A & B \\ 
C & D
\end{array}
\right) \text{, }  \label{Mblock}
\end{equation}
with
\begin{eqnarray*}
A &=&\left( \widehat{{\bf q}}\cdot \varepsilon \widehat{{\bf q}}\right) ^{-1}%
\widehat{q}^{\times }\varepsilon \widehat{{\bf q}}\circ {\bf a}-\left( 
\widehat{{\bf q}}\cdot \mu \widehat{{\bf q}}\right) ^{-1}{\bf b}\circ 
\widehat{{\bf q}}\mu I\text{, } \\
B &=&\left( \widehat{{\bf q}}\cdot \varepsilon \widehat{{\bf q}}\right)
^{-1}I\widetilde{\overline{\varepsilon }}I-\left( \widehat{{\bf q}}\cdot \mu 
\widehat{{\bf q}}\right) ^{-1}{\bf b}\circ {\bf b}\text{,} \\
\text{ }C &=&-\left( \widehat{{\bf q}}\cdot \varepsilon \widehat{{\bf q}}%
\right) ^{-1}{\bf a}\circ {\bf a}-\left( \widehat{{\bf q}}\cdot \mu \widehat{%
{\bf q}}\right) ^{-1}\widehat{q}^{\times }\widetilde{\overline{\mu }}%
\widehat{q}^{\times }\text{,} \\
D &=&\left( \widehat{{\bf q}}\cdot \varepsilon \widehat{{\bf q}}\right) ^{-1}%
{\bf a}\circ {\bf q}\varepsilon \widehat{q}^{\times }-\left( \widehat{{\bf q}%
}\cdot \mu \widehat{{\bf q}}\right) ^{-1}I\mu \widehat{{\bf q}}\circ {\bf b}%
\text{.}
\end{eqnarray*}
{\bf \ }Here $\varepsilon =1+\chi $ is the dielectric tensor, $v^{\times }$
denotes the dual tensor of an arbitrary vector ${\bf v}$ and the tilde sign
stands for the transpose of a tensor. The $I=-\left( \widehat{q}^{\times
}\right) ^2$operator projects a vector into the plane of the sample surface.
The tangential component of the incident wave vector is ${\bf b=}I\,{\bf k/}%
k $, and ${\bf a}:={\bf b}\times \widehat{{\bf q}}$ is a vector
perpendicular to the reflection plane, $\overline{\varepsilon }=\det \left(
\varepsilon \right) \varepsilon ^{-1}$, $\overline{\mu }=\det \left( \mu
\right) \mu ^{-1}$. Strictly speaking,{\bf \ }$A$, $B$, $C$ and $D$ are
3-dimensional tensors acting only, as it can be seen, in the ${\bf a,b}$
plane. Consequently, $M$ can be properly represented by $4\times 4$
matrices. The permeability tensor $\mu $ will play no further role.\cite
{Andreeva1} The solution of Eq. (\ref{BorzEq}) relates ${\bf H}_t\;$ and\ $%
\widehat{{\bf q}}\times {\bf E}$ to each other at the lower and upper
surfaces of the layered medium. In a homogeneous film of thickness $d$, the
solution is given by the so-called integral propagation matrix $L=\exp
\left( ikdM\right) $, by the matrix exponential of the differential
propagation matrix. For an $n$-layer system, the total integral propagation
matrix is the product of the individual integral propagation matrices $%
L_{\left( l\right) }$ of layer $l$, thus
\begin{equation}
L=L_{\left( n\right) }...L_{\left( 2\right) }L_{\left( 1\right) }\text{.}
\label{MultiL}
\end{equation}
The expression of the planar reflectivity, $r$ defined by ${\bf H}_t^r=r\/%
{\bf H}_t^0$, where ${\bf H}_t^r\;$ and\ ${\bf H}_t^0$ are the tangential
amplitudes of the reflected and incident waves, respectively, writes as
\begin{eqnarray}
r &=&\left( \left( \gamma ^t,-I_2\right) L\left( 
\begin{array}{c}
I_2 \\ 
\gamma ^r
\end{array}
\right) \right) ^{-1}  \nonumber \\
&&\times \left( \left( -\gamma ^t,I_2\right) L\left( 
\begin{array}{c}
I_2 \\ 
\gamma ^0
\end{array}
\right) \right) \text{.}  \label{Refl}
\end{eqnarray}
Here $I_2$ is the $2\times 2$ unity matrix and the $\gamma ^0$, $\gamma ^r$
and $\gamma ^t$ tensors are the impedance tensors for the incident,
specularly reflected and transmitted waves, respectively, defined by the
\begin{equation}
\gamma ^{0,r,t}{\bf H}_t^{0,r,t}=\widehat{{\bf q}}\times {\bf E}^{0,r,t}
\label{Gamma}
\end{equation}
equation. Since the $\gamma $ tensors act in the plane perpendicular to $%
\widehat{{\bf q}}$, they can be represented by $2\times 2$ matrices.\cite
{Borzdov}

The elements of the reflectivity matrix $R$ are geometrically related to the
elements of the planar reflectivity, $r$ i.e.
\begin{equation}
\begin{tabular}{ll}
$R_{\sigma \sigma }=-r_{22}$ & $R_{\sigma \pi }=-r_{21}\sin ^{-1}\theta $ \\ 
$R_{\pi \sigma }=r_{12}\sin \theta $ & $R_{\pi \pi }=r_{11}$%
\end{tabular}
\text{,}  \label{Trefl}
\end{equation}
where $\sigma $ and $\pi $ are polarizations corresponding to ${\bf E}$
perpendicular and parallel to the plane of incidence, respectively. For
numerical calculations we shall choose different appropriate coordinate
systems. The laboratory system ${\cal S}$ will be defined so that the $x,y\,$%
and $\,z$-axes are parallel to ${\bf a}$, ${\bf b}$ and $\widehat{{\bf q}}$,
respectively. The field components in Eq. (\ref{BorzEq}) define a natural
permutation ${\cal K}=\left( 1,2,3,4\right) $ basis of the $4$-component
field vectors $\left[ {\bf H}_x,{\bf H}_y,\left( \widehat{{\bf q}}\times 
{\bf E}\right) _x,\left( \widehat{{\bf q}}\times {\bf E}\right) _y\right] $
with respect to the ${\cal S}$-system. A convenient permutation of ${\cal K}%
=\left( 1,2,3,4\right) $ {\it viz}. ${\cal K}^{^{\prime }}=\left(
2,3,4,1\right) $ shall also be used. The differential and integral
propagation matrices $M$ and $L$ will be denoted by $M^{^{\prime }}$ and $%
L^{^{\prime }}$, respectively in the ${\cal K}^{^{\prime }}$-system.

\subsection{The forward scattering case: The Blume--Kistner equation}

\label{blumekistner}

The (\ref{Mblock}) differential propagation matrix for the case of normal
incidence\cite{Blume} in the ${\cal K}$-system has the following simple form
of
\begin{equation}
M=\left( 
\begin{array}{cccc}
0 & 0 & \varepsilon _{22} & -\varepsilon _{21} \\ 
0 & 0 & -\varepsilon _{12} & \varepsilon _{11} \\ 
1 & 0 & 0 & 0 \\ 
0 & 1 & 0 & 0
\end{array}
\right) \text{.}  \label{Mblume}
\end{equation}

The transmissivity may be expressed in terms of the integral propagation
matrix, $L=\exp \left( ikdM\right) $ as\cite{Borzdov}
\begin{equation}
t=2\left[ \left( I_2\,,I_2\right) L^{-1}\left( 
\begin{array}{c}
I_2 \\ 
I_2
\end{array}
\right) \right] ^{-1}  \label{Dforward}
\end{equation}
defined by ${\bf H}_t^t=t{\bf H}_t^0$ can be explicitly elaborated to obtain
the Blume--Kistner formulae.\cite{Blume} Indeed, using the identity (which
can easily be proved by expanding the exponentials):
\begin{widetext}
\begin{equation}
\exp \left( 
\begin{array}{cc}
0_2 & B \\ 
C & 0_2
\end{array}
\right) =\left( 
\begin{array}{cc}
\cosh \left( BC\right) ^{1/2} & B\left( CB\right) ^{-1/2}\sinh \left(
CB\right) ^{1/2} \\ 
C\left( BC\right) ^{-1/2}\sinh \left( BC\right) ^{1/2} & \cosh \left(
CB\right) ^{1/2}
\end{array}
\right) \text{,}  \label{Theorema}
\end{equation}
with $0_2$ being the $2\times 2$ zero matrix, the matrix exponential of $M$
can be expressed in terms of a $2\times 2$ submatrix $B=\left( 
\begin{array}{cc}
\varepsilon _{22} & -\varepsilon _{21} \\ 
-\varepsilon _{12} & \varepsilon _{11}
\end{array}
\right) $ of Eq. (\ref{Mblume}) so that the (\ref{Dforward}) transmissivity
\begin{equation}
t=\left[ \cosh \left( ikdB^{1/2}\right) -\frac 12\sinh \left(
ikdB^{1/2}\right) \left( B^{1/2}+B^{-1/2}\right) \right] ^{-1}\text{.}
\label{Tforw}
\end{equation}
\end{widetext}
Making use of the smallness of the susceptibility one can easily
write \thinspace $B^{1/2}+B^{-1/2}\approx 2I_2$ and the transmissivity:
\begin{equation}
t\approx \exp \left( ikd\,B^{1/2}\right) \text{.}  \label{Dpre}
\end{equation}
In order to compare the result (\ref{Dpre}) with those of Blume and Kistner,%
\cite{Blume} now we define the transmission coefficient for the electric
field by ${\bf E}^t=t_E\,{\bf E}^0$. Expressing ${\bf H}$ with ${\bf E}$ we
obtain the Blume--Kistner equation\cite{Blume}
\begin{equation}
t_E=-\widehat{q}^{\times }t\,\widehat{q}^{\times }\approx \exp \left(
ikd\,n\right) \text{,}  \label{Dfinal}
\end{equation}
where $n=\sqrt{1+\chi }$. Comparing Eq. (\ref{fDeak}) with Eq. (\ref{Dfinal}%
) we obtain the Lax formula\cite{Lax} as generalized by Blume and Kistner%
\cite{Blume}
\begin{equation}
\,n_{pp^{^{\prime }}}=\delta _{pp^{^{\prime }}}+\frac{2\pi \/N}{k^2}f^{{\bf k%
},p\rightarrow {\bf k},p^{^{\prime }}}\text{,}  \label{Laxform}
\end{equation}
where $\delta $ is the Kronecker symbol and $p,p^{^{\prime }}=\sigma ,\pi $.

\subsection{The grazing incidence case: The Andreeva approximation}

\subsubsection{The differential propagation matrix}

In order to see which elements in Eq. (\ref{Trefl}) are of the same order of
magnitude, we eliminate the explicit $\theta $-dependence of $R$ by applying
a linear transformation $T$ (in the ${\cal K}^{^{\prime }}$-system) of the
form: 
\begin{equation}
T=\left( 
\begin{array}{cccc}
\sin ^{-1}\theta & 0 & 0 & 0 \\ 
0 & \sin ^{-1}\theta & 0 & 0 \\ 
0 & 0 & 1 & 0 \\ 
0 & 0 & 0 & 1
\end{array}
\right) \text{.}  \label{T}
\end{equation}
It can be easily seen that only the integral propagation matrix
\begin{equation}
L_{(l)}^{^{\prime \prime }}=TL_{\left( l\right) }^{^{\prime }}T^{-1}
\label{L"}
\end{equation}
depends on $\theta $, and the reflectivity matrix depends on the elements of 
$L_{\left( l\right) }^{^{\prime \prime }}$ only. The transform of the
differential propagation matrix $M_{(l)}^{^{\prime \prime
}}=TM_{(l)}^{^{\prime }}T^{-1}$ of layer $l$ is obtained with the same
similarity transformation:
\begin{widetext}
\begin{equation}
M_{(l)}^{^{\prime \prime }}=\sin \theta \left( 
\begin{array}{cccc}
0 & 0 & 1 & 0 \\ 
0 & 0 & 0 & 1 \\ 
1 & 0 & 0 & 0 \\ 
0 & 1 & 0 & 0
\end{array}
\right) +\frac 1{\sin \theta }\left( 
\begin{array}{cccc}
0 & 0 & \chi _{(l)11} & \chi _{(l)13} \\ 
0 & 0 & \chi _{(l)31} & \chi _{(l)33} \\ 
0 & 0 & 0 & 0 \\ 
0 & 0 & 0 & 0
\end{array}
\right) +\left( 
\begin{array}{cccc}
0 & -\chi _{(l)12} & 0 & 0 \\ 
0 & -\chi _{(l)32} & 0 & 0 \\ 
0 & 0 & 0 & 0 \\ 
0 & \chi _{(l)22}\sin \theta & -\chi _{(l)21} & -\chi _{(l)23}
\end{array}
\right) \text{,}  \label{GrazM}
\end{equation}
$\chi _{(l)ij}$ $(i,j=1,2,3)$ being the matrix elements of the
susceptibility tensor of layer $l$.

The three matrices in Eq. (\ref{GrazM}) are of the order of magnitude of $%
\theta $, $\chi /\theta $ and $\chi $, respectively. Without a rigorous
explanation, Andreeva {\it et al.}\cite{Andreeva3} intuitively drop the
third term containing only those elements of the $\chi $ tensor which are
not related to the forward scattering amplitude. This approximation is
obviously valid if $\chi $ is small compared to $\theta $ and $\chi /\theta $%
. Since typically $\chi \approx 10^{-5}$ the interval for $\theta $ in order
the third term to remain below $1\%$ of the first two is: $10^{-3}<\theta
\leq 10^{-2}$ which is, indeed, the typical region of a grazing incidence
experiment. From the (\ref{GrazM}) form of $M$ it is clearly seen what
conditions have to be fulfilled for the Andreeva approximation to be valid.
Note that there is not only an upper but also a lower bound for $\theta $ .

Returning to the covariant notation, the differential propagation matrix $%
M_{(l)}$ of layer $l$ with a $1\%$ accuracy in the ${\cal K}$-system is of
the form
\begin{equation}
M_{\left( l\right) }=\left( 
\begin{array}{cc}
\left( {\bf a\cdot }\chi _{\left( l\right) }\widehat{{\bf q}}\right) {\bf b}%
\circ {\bf a} & I-{\bf b}\circ {\bf b}\left[ 1-\left( {\bf a\cdot }\chi
_{\left( l\right) }{\bf a}\right) \right] \\ 
I-{\bf a}\circ {\bf a}\left[ 1-\left( \widehat{{\bf q}}\cdot \chi _{\left(
l\right) }\widehat{{\bf q}}\right) \right] & \left( \widehat{{\bf q}}\cdot
\chi _{\left( l\right) }{\bf a}\right) {\bf a}\circ {\bf b}
\end{array}
\right)  \label{Andre}
\end{equation}
which is identical to the form suggested by Andreeva {\it et al.}\cite
{Andreeva3}. In the grazing incidence case the ${\bf a\,}$and${\bf \,b}$
vectors are approximate unit vectors $\left| {\bf a}\right| =\left| {\bf b}%
\right| =\cos \theta \approx 1$. This approximation is equivalent to
neglecting terms of the order of $\sin ^2\theta $ as compared to $1$. In
this limit ${\bf k\Vert b}${\bf .} We can choose the two polarization
vectors so that $\widehat{{\bf u}}_1\approx {\bf a}$ and $\widehat{{\bf u}}%
_2\approx \widehat{{\bf q}}$. Transforming the $\chi $ matrix given in the
polarization vector system ${\cal P}$ by Eq. (\ref{fDeak}) into the ${\cal S}
$ system and substituting into Eq. (\ref{Andre}) the differential
propagation matrix can be expressed in terms of the forward scattering
amplitude:
\begin{equation}
M_{\left( l\right) }\approx \left( 
\begin{array}{cccc}
0 & \,\,\,\,0\,\,\,\, & \,\,\,\,1\,\,\,\, & 0 \\ 
\frac{4\pi N_{\left( l\right) }}{k^2}f_{\left( l\right) }^{\ {\bf k},\sigma
\rightarrow {\bf k},\pi } & 0 & 0 & \frac{4\pi N_{\left( l\right) }}{k^2}%
f_{\left( l\right) }^{{\bf k},\sigma \rightarrow {\bf k},\sigma }+\sin
^2\theta \\ 
\frac{4\pi N_{\left( l\right) }}{k^2}f_{\left( l\right) }^{{\bf k},\pi
\rightarrow {\bf k},\pi }+\sin ^2\theta & 0 & 0 & \frac{4\pi N_{\left(
l\right) }}{k^2}f_{\left( l\right) }^{{\bf k},\pi \rightarrow {\bf k},\sigma
} \\ 
0 & 1 & 0 & 0
\end{array}
\right) \text{.}  \label{deak}
\end{equation}
\end{widetext}
which, as we shall see, is a particularly suitable form for
numerical calculations ($N_{\left( l\right) }$ is the number of resonant
nuclei per unit volume and $f_{\left( l\right) }^{{\bf k},p\rightarrow {\bf k%
},p^{^{\prime }}}$ is the coherent forward scattering amplitude in layer $l$%
). Starting with Eq. (\ref{deak}) a time-effective numerical algorithm is
derived in the following sub-section.

\subsubsection{Numerical calculations}

The matrix (\ref{deak}) contains small quantities of the order of $\sin
^2\theta <$ $10^{-4}$ and the much larger number unity. The calculation of
the exponential of $M_{\left( l\right) }$ to a sufficient accuracy is rather
time consuming even if the approximation $\exp y\approx (1+\frac y{2^n}%
)^{2^n}$ is applied. For each energy channel the exponential of $M$ should
be calculated thus typically $2^{10}$ times per M\"{o}ssbauer spectrum.

The corresponding transformed integral propagation matrix in the ${\cal K}%
^{^{\prime }}$-system [cf. Eq. (\ref{GrazM})] can be written as 
\begin{eqnarray}
L_{\left( l\right) }^{^{\prime \prime }} &=&\exp \left( ikd_{\left( l\right)
}TM_{\left( l\right) }^{^{\prime }}T^{-1}\right)  \nonumber \\
&=&\exp \left( 
\begin{array}{cc}
0_2 & x_{\left( l\right) }I_2+\frac 1{x_{\left( l\right) }}\phi _{\left(
l\right) } \\ 
x_{\left( l\right) }I_2 & 0_2
\end{array}
\right) \text{,}  \label{Spiform}
\end{eqnarray}
where $x_{\left( l\right) }=ikd_{\left( l\right) }\sin \theta $, with
$d_{\left( l\right) }$ being the thickness of layer $l$. $\phi _{\left(
l\right) }=-4\pi N_{\left( l\right) }d_{\left( l\right) }^2f_{\left(
l\right) }$ is proportional to the forward scattering amplitude 
$f_{\left(l\right) }$.

To evaluate the integral propagation matrix (\ref{Spiform}) one may notice
that the differential propagation matrix is block-anti-diagonal. We show
that the problem, like in the Blume--Kistner case in Sec. \ref{blumekistner}
reduces to the calculation of a single $2\times 2$ matrix exponential of a
small quantity. Indeed, using again the identity (\ref{Theorema}), with $%
B_{\left( l\right) }=x_{\left( l\right) }I_2+\frac 1{x_{\left( l\right) }}%
\phi _{\left( l\right) }$ and $C_{\left( l\right) }=x_{\left( l\right) }I_2$
the integral propagation matrix of Eq. (\ref{Spiform}) with $F_{\left(
l\right) }=$ $\left( x_{\left( l\right) }B_{\left( l\right) }\right) ^{1/2}$
is given by:
\begin{equation}
L_{\left( l\right) }^{^{\prime \prime }}=\left( 
\begin{array}{cc}
\cosh F_{\left( l\right) } & \frac 1{x_{\left( l\right) }}F_{\left( l\right)
}\sinh F_{\left( l\right) } \\ 
x_{\left( l\right) }F_{\left( l\right) }^{-1}\sinh F_{\left( l\right) } & 
\cosh F_{\left( l\right) }
\end{array}
\right) \text{.}  \label{Final}
\end{equation}
Eq. (\ref{Final}) is well suited for numerical calculations since it
contains only the $2\times 2$ matrix exponential $\exp F_{\left( l\right) }$.

By the present method the large matrix elements are separated from the small
ones. If the argument of the exponential is of the order of $10^{-4}$ the $%
\exp y\approx (1+\frac y{2^n})^{2^n}$ approximation gives a sufficient
accuracy with $n$ as small as $2$.

Using Eqs. (\ref{Refl}),(\ref{Trefl}),(\ref{MultiL}), and (\ref{Final}) the
reflectivity in the $\sigma ,\pi $ basis is given by
\begin{widetext}
\begin{equation}
R=\left( L_{\left[ 11\right] }^{^{\prime \prime }}-L_{\left[ 12\right]
}^{^{\prime \prime }}-L_{\left[ 21\right] }^{^{\prime \prime }}+L_{\left[
22\right] }^{^{\prime \prime }}\right) ^{-1}\left( L_{\left[ 11\right]
}^{^{\prime \prime }}+L_{\left[ 12\right] }^{^{\prime \prime }}-L_{\left[
21\right] }^{^{\prime \prime }}-L_{\left[ 22\right] }^{^{\prime \prime
}}\right) \text{,}  \label{Rfinal}
\end{equation}
\end{widetext}
where the $L_{\left[ ab\right] }^{^{\prime \prime }}$s ($a,b=1,2$) are $%
2\times 2$ submatrices of the integral propagation matrix $L^{^{\prime
\prime }}$ (see Appendix). Since Eq.(\ref{Rfinal}) gives the reflected
amplitude rather than the reflected intensity it is equally applicable in
calculating spectra in conventional source ({\it i.e.} energy domain) and in
synchrotron radiation ({\it i.e.} time domain) experiments. Using the
present method a computer program was developed capable of {\it fitting}
experimental spectra both in the energy and in the time domain.

\section{SUMMARY}

The goal of the present paper was twofold. First, to establish a working
theory of M\"{o}ssbauer spectroscopy by specularly reflected $\gamma $-rays
for both the conventional source and for the synchrotron radiation
experiment, and second, deriving the corresponding formulae in a numerically
tractable form. Starting from the nucleon current density expression of the
susceptibility tensor of Afanas'ev and Kagan\cite{Kagan} we use a covariant
formalism\cite{Borzdov} of anisotropic optics. Both in the transmission and
in the grazing incidence geometry the susceptibility is expressed in terms
of the coherent forward scattering amplitude. The Blume--Kistner formula\cite
{Blume} of the perpendicular transmissivity and the Andreeva approximation%
\cite{Andreeva3} for the grazing incidence reflectivity are re-derived in a
rigorous manner. In the grazing incidence case a concise $2\times 2$
block-matrix exponential expression for the differential propagation matrix
is obtained for transitions of arbitrary multipolarity and in a numerically
convenient way. This latter allows for fast numerical calculation and
practical fitting of M\"{o}ssbauer spectra both in energy and in time domain.

\section*{ACKNOWLEDGMENT}

Fruitful discussions with Dr. M. A. Andreeva are gratefully acknowledged.
This work was partly supported by the PHARE ACCORD Program under Contract
No. H--9112--0522 and by the Hungarian Scientific Research Fund (OTKA) under
Contract Nos. 1809 and T016667. The authors also thank for the partial
support by the Deutsche Forschunsgemeinschaft and the Hungarian Academy of
Sciences in frames of a bilateral project.

\appendix

\section*{Derivation of the reflectivity formula}
\label{app}

The integral propagation matrix $L$ of Eq. (\ref{Refl}) is expressed by $%
L^{^{\prime \prime }}$ of Eq. (\ref{L"}):
\begin{eqnarray}
r &=&\left( \left( \gamma ^t,-I_2\right) V^{-1}T^{-1}L^{\prime \prime
}TV\left( 
\begin{array}{c}
I_2 \\ 
\gamma ^r
\end{array}
\right) \right) ^{-1}  \nonumber \\
&&\times \left( \left( -\gamma ^t,I_2\right) V^{-1}T^{-1}L^{\prime \prime
}TV\left( 
\begin{array}{c}
I_2 \\ 
\gamma ^0
\end{array}
\right) \right) ,  \label{ARefl}
\end{eqnarray}
where $T$ is given in Eq. (\ref{T}), $\gamma ^{t,r,0}$ are the impedance
tensors for the transmitted, reflected and incident radiation as defined in
Eq. (\ref{Gamma}). Assuming vacuum on both sides of the stratified sample
(which --- by allowing for a thick enough substrate --- imposes no further
restriction) the $\gamma $ 's are of the form \cite{Borzdov} 
\begin{equation}
\gamma ^0=\gamma ^t=-\gamma ^r=\left( 
\begin{array}{cc}
\sin \theta & 0 \\ 
0 & \sin ^{-1}\theta
\end{array}
\right)  \label{Gammavac}
\end{equation}
$V$ is the matrix of the $\left( 1,2,3,4\right) $ {\it viz}. $\left(
2,3,4,1\right) $ (${\cal K}\longrightarrow {\cal K}^{^{\prime }}$)
transformation of the form:
\begin{equation}
V=\left( 
\begin{array}{cccc}
0 & 1 & 0 & 0 \\ 
0 & 0 & 1 & 0 \\ 
0 & 0 & 0 & 1 \\ 
1 & 0 & 0 & 0
\end{array}
\right)  \label{V}
\end{equation}
Performing the calculations in Eq. (\ref{ARefl}) with the above matrices the
planar reflectivity:
\begin{eqnarray}
r &=&-\left( 
\begin{array}{cc}
0 & \sin ^{-1}\theta \\ 
-1 & 0
\end{array}
\right) ^{-1}\left( L_{\left[ 11\right] }^{^{\prime \prime }}-L_{\left[
12\right] }^{^{\prime \prime }}-L_{\left[ 21\right] }^{^{\prime \prime
}}+L_{\left[ 22\right] }^{^{\prime \prime }}\right) ^{-1}  \nonumber \\
&&\times \left( L_{\left[ 11\right] }^{^{\prime \prime }}+L_{\left[
12\right] }^{^{\prime \prime }}-L_{\left[ 21\right] }^{^{\prime \prime
}}-L_{\left[ 22\right] }^{^{\prime \prime }}\right) \left( 
\begin{array}{cc}
0 & \sin ^{-1}\theta \\ 
1 & 0
\end{array}
\right)  \label{Apr}
\end{eqnarray}

From Eqs. (\ref{Apr}) and (\ref{Trefl}) we obtain the (\ref{Rfinal})
reflectivity formula.

%


\begin{thebibliography}{35}%
\bibitem{Lax}  M. Lax, Rev. Mod. Phys. {\bf 23}, 287 (1951).

\bibitem{Blume}  M. Blume and O. C. Kistner, Phys. Rev. {\bf 171}, 417 (1968)

\bibitem{Frost}  J. C. Frost, B. C. C. Cowie, S. N. Chapman, and J. F.
Marshall, Appl. Phys. Lett. {\bf 47}, 581 (1985).

\bibitem{Nagy}  D. L. Nagy and V. V. Pasyuk, Hyp. Int. {\bf 71}, 1349 (1992)

\bibitem{Irkaev1}  S. M. Irkaev, M. A. Andreeva, V. G. Semenov, G. N.
Beloserskii, and O. V. Grishin, Nucl. Instrum. Methods {\bf B74}, 545 (1993).

\bibitem{Irkaev2}  S. M. Irkaev, M. A. Andreeva, V. G. Semenov, G. N.
Beloserskii, and O. V. Grishin, Nucl. Instrum. Methods {\bf B74}, 554 (1993).

\bibitem{Andreeva2}  M. A. Andreeva, S. M. Irkaev, and V. G. Semenov, Sov.
Phys.--JETP {\bf 78}, 965 (1994).

\bibitem{Andreeva1}  M. A. Andreeva and R. N. Kuz'min, {\it Messbauerovskaya
Gamma-Optika} (Moscow University, 1982).

\bibitem{Andreeva3}  M. A. Andreeva and K. Rosete, Poverkhnost' {\bf 9}, 145
(1986); Vestnik Mosk. Univ., Ser. 3. Fiz. Astron. {\bf 27}, 57 (1986).

\bibitem{Kagan}  A. M. Afanas'ev and Yu. Kagan, Sov. Phys.--JETP {\bf 21},
215 (1965).

\bibitem{Borzdov}  G. M. Borzdov, L. M. Barskovskii, and V. I. Lavrukovich,
Zh. Prikl. Spektrosk. {\bf 25}, 526 (1976).

\bibitem{Fedorov}  F. I. Fedorov, {\it Teoria Girotropii} (Nauka i Technika,
Minsk, 1976).

\bibitem{Hannon1}  J. P. Hannon and G. T. Trammell, Phys. Rev. {\bf 169},
315 (1968).

\bibitem{Hannon2}  J. P. Hannon and G. T. Trammell, Phys. Rev. {\bf 186},
306 (1969).

\bibitem{Hannon3}  J. P. Hannon, N. V. Hung, G. T. Trammell, E. Gerdau, M.
Mueller, R. R\"{u}ffer, and H. Winkler, Phys. Rev. B{\bf \ 32}, 5068 (1984).

\bibitem{Hannon4}  J. P. Hannon, G. T. Trammell, M. Mueller, E. Gerdau, R.
R\"{u}ffer, and H. Winkler, Phys. Rev. B{\bf \ 32}, 6363 (1985).

\bibitem{Roehlsberger}  R. R\"{o}hlsberger, Ph.D. thesis, University
Hamburg, 1994.

\bibitem{Spiering85}  H. Spiering, Hyp. Int. {\bf 24--26}, 737 (1985).
\end{thebibliography}
\end{document}